\documentclass[%
 reprint,
 amsmath,amssymb,
 aps,
]{revtex4-1}

\usepackage{graphicx}
\usepackage{dcolumn}
\usepackage{bm}
\usepackage{color}
\usepackage{soul}

\usepackage{amsmath}
\usepackage{amssymb}
\usepackage{latexsym}
\usepackage{bm}
\usepackage{color}
\usepackage{upgreek}
\usepackage{dcolumn}
\newcommand{\Fr}{{\rm Fr}}
\newcommand{\Bo}{{\rm Bo}}
\usepackage{soul}
\newcommand{\be}{\begin{equation}}
\newcommand{\ee}{\end{equation}}
\newcommand{\bea}{\begin{eqnarray}}
\newcommand{\eea}{\end{eqnarray}}

\begin{document}

\preprint{APS/123-QED}

\title{Leidenfrost droplets trampoline}

\author{Dongdong Liu\textit{$^{a}$}\textit{$^{b}$}}

\author{Tuan Tran\textit{$^{a}$}\textit{$^{b}$}}%
 \email{ttran@ntu.edu.sg}
\affiliation{%
 \textit{$^{a}$School of Mechanical$\&$Aerospace Engineering, Nanyang Technological University, 50 Nanyang Avenue, 639798, Singapore.}
 \textit{$^{b}$HP-NTU Digital Manufacturing Corporate Lab, Nanyang Technological University, 50 Nanyang Avenue, 639798 Singapore.}
}%

\date{\today}

\begin{abstract}
A liquid droplet hovering on a hot surface is commonly referred to as a Leidenfrost droplet.
In this study, we discover that a Leidenfrost droplet involuntarily
performs a series of distinct oscillations
as it shrinks during the span of its life.
The oscillation first starts out erratically, followed by a stage with stable frequencies,
and finally turns into periodic bouncing with signatures of a parametric oscillation and occasional resonances.
The last bouncing stage exhibits nearly perfect collisions.
We showed experimentally and theoretically the enabling effects of each oscillation mode and how the droplet switches between such modes.
We finally show that these self-regulating oscillation modes and the conditions for transitioning between modes are universal for all
tested combinations of liquids and surfaces.
\end{abstract}

\pacs{Valid PACS appear here}
\maketitle
Leidenfrost effect, a two-century old phenomenon \cite{leidenfrost1966fixation,quere2013leidenfrost}
causing levitation of liquid droplets
deposited on hot surfaces,
has been playing a critical role in an increasing number of modern technologies.
As the effect completely removes
liquid-solid contact and subsequently liberates the liquid
from  frictional constraint of the surrounding,
it has great potential to transform liquid-transport applications
ranging from largescale drag reduction \cite{vakarelski2016leidenfrost},
rapid and autonomous transport of liquid droplets \cite{linke2006self,li2016directional},
to nanoscale manufacturing processes \cite{cordeiro2016leidenfrost}.
Our current understanding of the Leidenfrost phenomenon
is largely based on the steady-state assumption, an approach
used to justify exclusion of minute but accumulative effects
such as drop size reduction by evaporation.
The resulting analysis, while offers tremendous insights into the
short-time Leidenfrost dynamics,
filters out phenomena only visible
at longer timescales, e.g., the life time of Leidenfrost droplets.

\begin{figure}[t!]
	\centering
	\includegraphics[width=72mm]{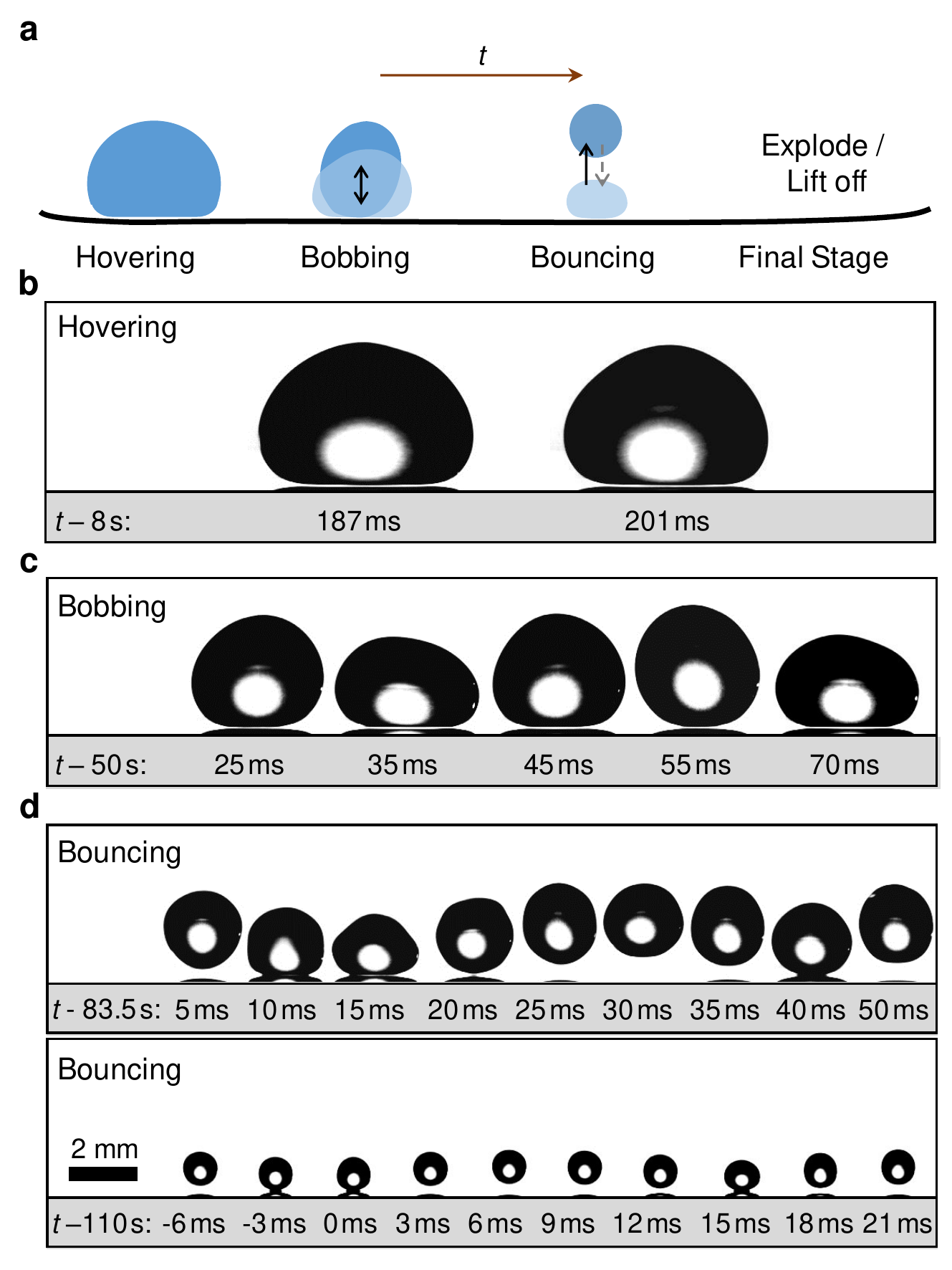}
	\caption{(a) Schematic illustrating the characteristic behaviors of a
		Leidenfrost droplet as it evaporates and shrinks.
		The initial drop size typically is larger than the liquid's capillary length.
		Once deposited on a superheated surface
		the droplet goes through the following stages:
		(b) hovering (c) bobbing and (d) bouncing.
		Eventually, when the drop becomes sufficiently small, it
		reaches the final stage in which
		it either lifts off or explodes \cite{celestini2012take,lyu2019final}.
		The snapshots from panels (b) to (d) were taken from an experiment using
		DI water as the working fluid. The surface was polished
		aluminum surface and heated to 380\,$^{\circ}$C.}
	\label{fig1}
	\vskip -0.5cm
\end{figure}

Here,
we reveal that as a Leidenfrost droplet shrinks,
it involuntarily performs a series of trampolining motions,
starting erratically at the beginning,
followed by regular oscillation
and finally
settling at periodic bouncing towards the end of its life time.
The bouncing behavior has an unusually high restitution coefficient
and exhibits
signatures of parametric oscillation with occasional resonances.
Our findings demonstrate the active nature of Leidenfrost droplets at long timescales
by showing that they exhibit self-regulating ability by switching through different modes of oscillation
in response to reduction in drop size by evaporation.
The underlying active mechanism
may serve as a basis to explore
strategies for energy harvesting or frictionless and autonomous liquid transport.

\begin{figure*}[t!]
	\centering	\includegraphics[width=150mm]{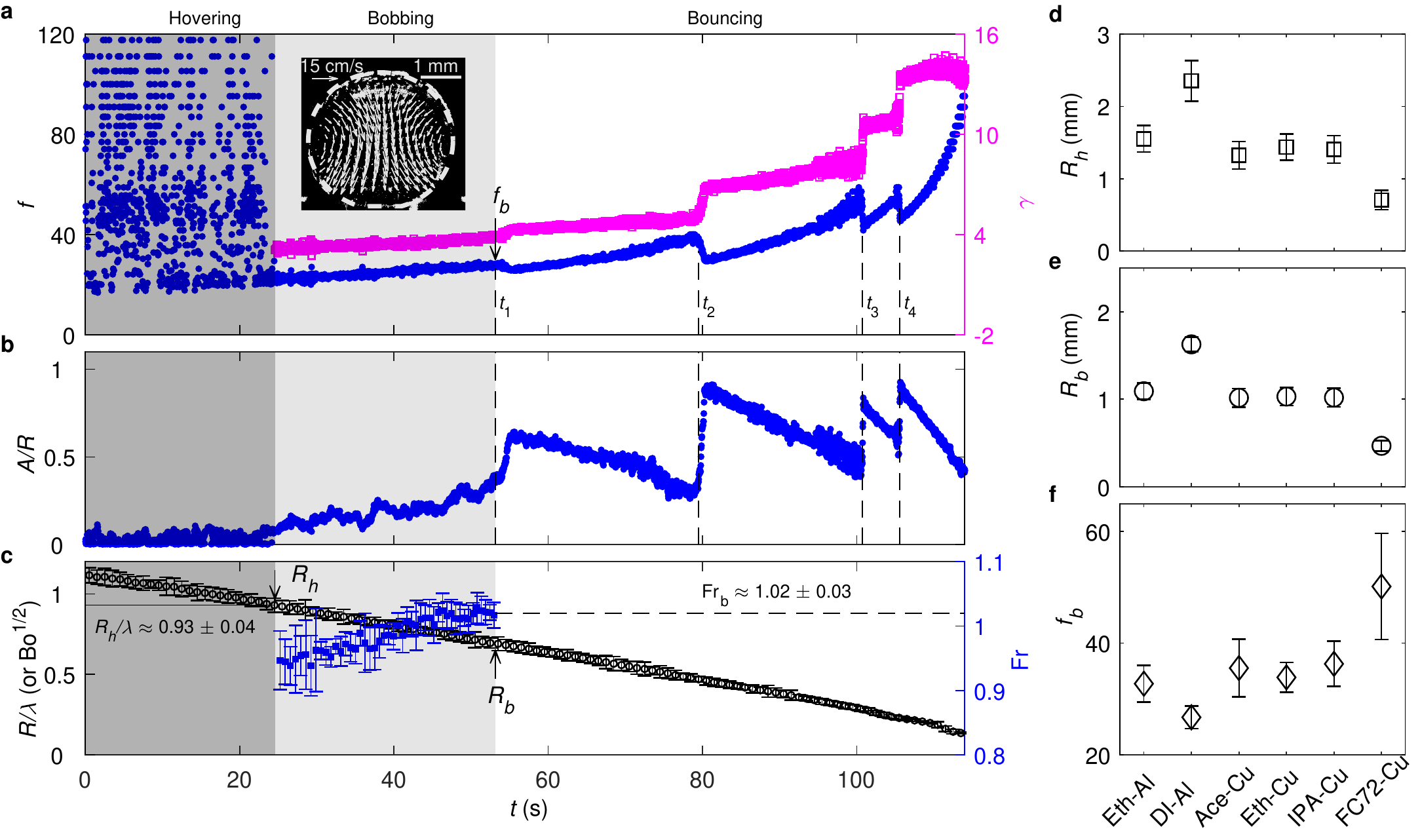}
	\caption{(a) Representative plot showing the oscillation frequency $f$ (left axis) and
		the ratio $\gamma=2f_n/f$ (right axis)  \emph{versus} time $t$
		as the droplet transitions through several characteristic stages: hovering, bobbing and bouncing.
		Parametric resonance occurs 
		in the bouncing stage at $t_1$, $t_2$, $t_3$ and $t_4$.
		Inset: Internal velocity field of a droplet in the bobbing stage.
		(b) Dimensionless oscillation amplitude $A/R$ \emph{versus} time $t$.
		(c) Dimensionless radius $R/\lambda = \Bo^{1/2}$ (left axis)
		and Froude number $\Fr$ in the bobbing stage (right axis) \emph{versus} $t$.
		The solid line indicates that the transition from
		hovering to bobbing stage happens at
		$R/\lambda \approx 0.93$,
		or equivalently $\Bo \approx 0.86$.
		The dashed line indicates that the
		transition from bobbing to bouncing
		occurs at $\Fr \approx1.0$.
		Variation in (d) droplet radius $R_h$
		at the hovering-bobbing transition
		(e) droplet radius $R_b$ and (f)
		frequency $f_b$ at the bobbing-bouncing transition
		for various fluid-surface combinations.
		The tested fluids are ethanol (Eth), DI water (DI), acetone (Ace),
		IPA and FC72. The tested surface materials are
		copper (Cu) and aluminum (Al).}
		\label{fig2}
		\vskip -0.3cm
\end{figure*}

The spontaneous trampolining motion,
of a Leidenfrost droplet
is self-triggered and
occurs as the droplet loses its
weight to evaporation.
The transitions between characteristic behaviors
inevitably result from
liquid evaporation and subsequent size
reduction,
from
the liquid's capillary length
(several millimeters)
to hundreds of micrometers.
In considering the underline phenomena,
in particular the droplet's vertical motion,
we trap Leidenfrost droplets horizontally
by using a slightly curved smooth surface,
thereby minimizing the Leidenfrost wheeling effect
\cite{bouillant2018}
that triggers the horizontal motion.

During the lifetime of a Leidenfrost droplet,
its behavior transitions through
several stages,
as illustrated in
the schematic shown in Fig.~\ref{fig1}a.
A droplet 
larger than the liquid's capillary length
after deposited on a sufficiently heated surface
first hovers around on its own vapor layer.
In this so-called \emph{hovering} stage (Fig.~\ref{fig1}\,b),
the droplet has a relatively large
flatten area at the bottom
due to gravity and this area does not
vary significantly with time.
As the droplet gets smaller due to
evaporation,
capillary force becomes dominating over
gravitational force,
causing its behavior transition
from hovering to \emph{bobbing},
i.e., periodic
vertical deformation without leaving the surface
(Fig.~\ref{fig1}\,c).
When the droplet radius
reduces to a
critical value,
it starts \emph{bouncing} on the surface,
i.e.,
the droplet is no longer
separated from the surface by a thin
vapor film but jumps up and down periodically.
(Fig.~\ref{fig1}\,d).
Once the droplet becomes
sufficiently small
($R\sim100\,\mu$m),
it either takes a final leap
out of the camera's view
or explodes \cite{celestini2012take,lyu2019final}.

The recorded phenomenological behaviors
of Leidenfrost droplets
are robust for all tested liquids,
including deionized (DI) water,
ethanol, isopropyl alcohol (IPA), acetone and
FC-72,
as well as for surface materials
such as copper and aluminum
\cite{otherliquids}.
We also verify these behaviors
for a wide range of
surface temperature,
from 180\,$^{\circ}$C to 440\,$^{\circ}$C,
confirming that the phenomena are not
material and temperature specific.

To shed light into the observed
transient dynamics
of Leidenfrost droplets,
we track the
vertical position $y_c$
of the center of mass as a function of time.
Subsequently, the frequency $f$
and amplitude $A$ of the periodic motion of the center
of mass can be extracted.
In Fig.~\ref{fig2}a and b,
we show representative plots
of $f$ and the
normalized amplitude $A/R$
for a DI water droplet
on a aluminum surface at 380\,$^{\circ}$C.
Here $R$ is the droplet radius calculated by
assuming a spherical droplet of the same
volume with the one recorded from
the side view \cite{Bond}.
The frequency plot presents a
clear picture of the transition from the hovering
stage, where
large scattering in $f$ is observed,
to the bobbing stage, where $f$
gently increases with time.

The erratic oscillation of the droplet in the hovering stage
originates from capillary waves on its surface.
Indeed, the droplet radius
measured in this stage is larger than the
capillary length $\lambda=(\sigma /\rho g)^{1/2}$
(see Fig.~\ref{fig2}c).
Here, $\rho$ and $\sigma$ respectively are
the density and surface tension
at the liquid's boiling point
\cite{biance2003,bouillant2018}.
As the droplet radius
becomes smaller than $\lambda$
the capillary waves diminish, giving rise
to more regular oscillations.
As a result, we
conclude that the transition from hovering
to bobbing
occurs when $R \approx \lambda$,
or equivalently when the
Bond number
${\rm Bo} = \rho g R^2/\sigma = (R/\lambda)^2 \approx 1$.

We now focus on the
transition from bobbing to
bouncing.
Although not clearly displayed through $f$,
this transition is clearly determined
from the recording
when the droplet starts jumping readily from the surface.
We observe experimentally that
the internal flow of a droplet in the bobbing stage
resembles a toroidal field, i.e.,
a strong downward flow at the center of the droplet
(Fig.~\ref{fig2}a, inset) \cite{piv1}.
This flow field provides a crucial evidence indicating
the driving mechanism of the droplet's oscillation and the eventual transition to bouncing:
the temperature difference between the bottom and top
of the droplet generates a thermocapillary effect that
induces the internal flow, the associated pressure difference inside the droplet,
and subsequently
its surface deformation.
As the droplet radius continuously
decreases, the involving  %
parameters of oscillation,
i.e., the frequency associated with the internal flow
and the droplet's natural frequency,
also vary with time,
although at different rates
\cite{rate}.
This eventually leads the oscillating droplet
to parametric resonances and the observed
abrupt growths in its amplitude (Fig.~\ref{fig2}b).
The transition to bouncing therefore occurs
when the excited oscillation
gains sufficient upward acceleration
to overcomes gravity.
If we denote $V_i$ the
characteristic velocity of internal flow,
then the
acceleration associated with the internal flow ($a = V_i^2/2R$)
in comparison with the gravitational acceleration
is evaluated
by the
Froude number
${\rm Fr} = (a/g)^{1/2} = V_i/(2Rg)^{1/2}$.
In other words, the droplet overcomes gravity
when ${\rm Fr} \gtrsim 1$.


To test the hypothesis that
the transition from bobbing to bouncing
is possible at ${\rm Fr} \approx 1$,
we now examine the dependence of the Froude number
on the internal flow characteristics, in particular its frequency $f_i$.
Since the oscillation is driven by the internal flow
and recall that $f$ is the oscillation frequency of the droplet,
we have $f \approx f_i \approx V_i/4R$, giving $V_i \approx 4Rf$.
By substituting the expression of $V_i$ into the one for $\Fr$
and using the natural frequency $f_n$ of the droplet
to normalize $f$, we
obtain the following expression for $\Fr$:
\begin{equation}
{\rm Fr} \approx  \left (\frac{6}{\pi}\right )^{1/2}\frac{f}{f_n} \left(\frac{R}{\lambda} \right)^{-1}.
\end{equation}
Here, $f_n = (\sigma/m)^{1/2}$,
where $m=(4 \pi/3)\rho R^3$ is the mass of the droplet \cite{molavcek2013drops,gilet2009fluid,schutzius2015}.
In Fig.~\ref{fig2}c, we show how $\Fr$ changes
in the bobbing stage. Indeed,
the condition $\Fr \approx 1$
holds
at the transition to bouncing,
indicating that the upward acceleration caused by
internal flows overcomes the gravitational acceleration
at the transition from bobbing to bouncing.

\begin{figure}[t]
	\centering
	\includegraphics[width=57mm]{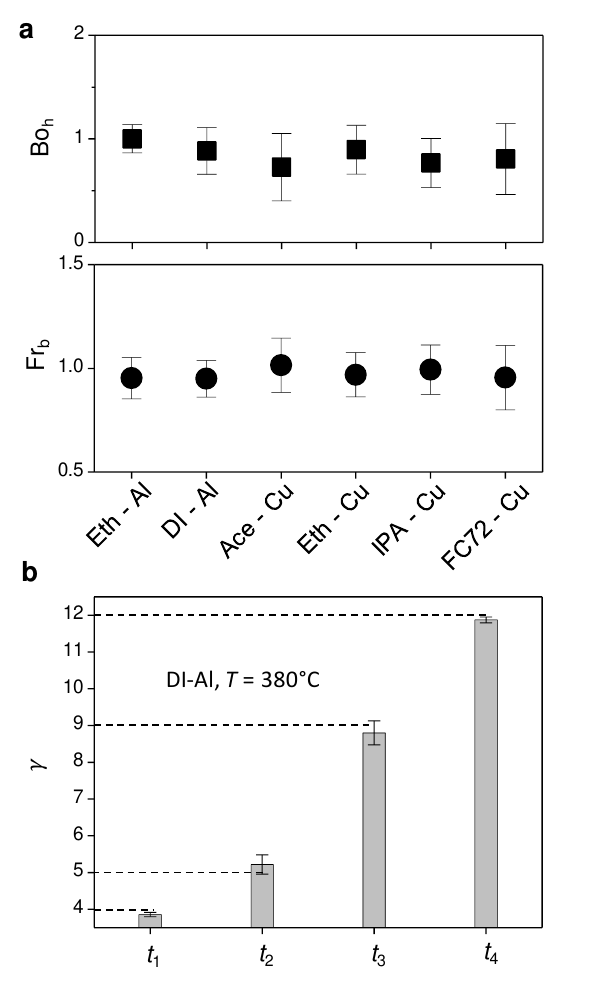}
	\caption{(a) Plots showing
		the Bond number ${\rm Bo}_{\rm h}$ at the hovering--bobbing transition
		and the Froude number ${\rm Fr}_{\rm b}$
		at the bobbing--bouncing transition for
		various fluid-surface combinations.
		(b) Plot showing the ratio $\gamma=2f_n/f$
		evaluated at the time of resonances.
		Dashed lines indicate integer values of $\gamma$.}
	\label{fig3}
	\vskip -0.5cm
\end{figure}

The two transitions that we observed, therefore,
can be understood using the
dimensionless numbers $\Bo$ and $\Fr$.
The first transition occurs
when $\Bo \approx 1$ and the second transition
occurs when  $\Fr \approx 1$.
We confirm experimentally that
although different fluid--surface combinations
yield disparate values for
the radius $R_h$ at the hovering-bobbing transition,
the radius $R_b$, or the frequency $f_b$
at the bobbing-bouncing transition (Fig.~\ref{fig2}d--f) \cite{temperature},
the conditions for transitions always
hold,
as shown in Fig.~\ref{fig3}\,a.

\begin{figure*}[t]
	\centering
	\includegraphics[width=14cm]{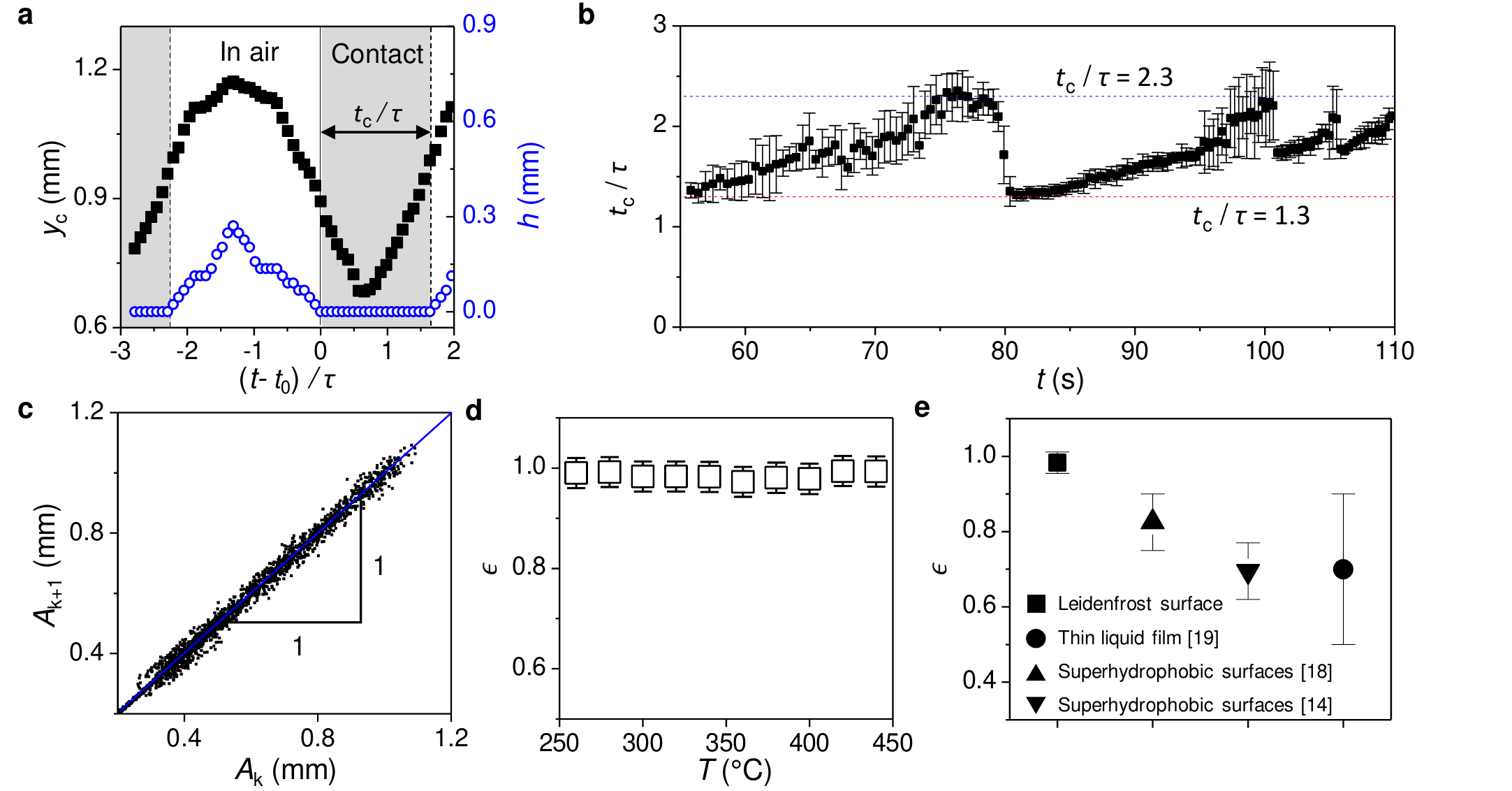}
	\caption{(a) Representative plots showing the
		vertical centre of mass $y_c$ (left axis, square markers)
		and the gap $h$ between the droplet and the surface (right axis, circular markers)
		\emph{versus}
		the dimensionless time $(t-t_0)/\tau$ in the bouncing stage.
		The shaded area indicates the contact time $t_c$ during which the droplet
		contacts with the surface ($h \approx 0$).
		(b) Dimensionless contact time $t_c/\tau$ \emph{versus}
		$t$ in bouncing stage.
		The dash lines represent the lower and upper bounds of $t_c/\tau$.
		(c) Amplitude ratio $A_{k+1}/A_{k}$ of two consecutive oscillations at
		any cycle $k$
		in the bouncing stage.
		(d) Restitution coefficient $\epsilon$ averaged over the entire bouncing stage
		of a water droplet \emph{versus} wall temperature $T$.
		(e) Comparison of restitution coefficient $\epsilon$ of water droplets on different surfaces:
		Leidenfrost surface (present study - squares),
		thin liquid film (circles),
		superhydrophobic surfaces (upward- and downward- triangles).
		}\label{fig4}
		\vskip -0.5cm
\end{figure*}

We now focus on the bouncing stage
to discuss the abrupt drops in $f$
and the corresponding jumps in $A$
(Fig.~\ref{fig2}a and b).
We note that
an increasing natural frequency
of the droplet,
combined with the
sudden drops in its frequency $f$
at $t_1$, $t_2$,
$t_3$ and $t_4$ (Fig.~\ref{fig2}a)
indicates that the droplet
may experience parametric resonances
at such moments
\cite{landau1976mechanics,ibrahim2008parametric}.
Indeed, if we
follow the signature of parametric
resonances and examine the
ratio $\gamma = 2f_n/f$,
we observe that
$\gamma$
increases substantially,
from $2.8$
at the bobbing-bouncing
transition to
$14.7$
at the end of the bouncing stage.
Whenever
there is a sudden drop in $f$, i.e.,
at $t_1$, $t_2$,
$t_3$, or $t_4$,
the corresponding value of $\gamma$
is approximately at the vicinity of an integer (see Fig.~\ref{fig3}b).
The first resonance at $t_1$ allows
the droplet to gain sufficient upward acceleration
to overcome gravitational acceleration and transition
to the bouncing stage.
As a result, we conclude that
the sudden decreases
in $f$ and the corresponding amplitude jumps
in the bouncing stage
result from
the droplet going through
parametric resonances
\cite{resonance}.


Between resonances,
the parametric oscillation of a bouncing droplet,
affected by its internal flow,
is accompanied by
a remarkable high \emph{contact time} \cite{bird2013reducing}
with the heated surface.
Here, we define the contact time $t_c$
of each
bouncing period using the
duration in which the droplet appears
in contact with the surface (see Fig.~\ref{fig4}a),
although strictly speaking, a Leidenfrost
droplet and the
heated surface are always separated by a
vapor layer.
In Fig.~\ref{fig4}b, we show the
variation of the normalized contact time $t_c/\tau$
in the bouncing stage, with
$\tau = 1/f_n$
being the natural bouncing period.
The plot shows that $t_c/\tau$
varies between 1.3 and 2.3,
and is mostly higher
than that in the
case of impacting droplet on unheated surface,
where $t_c/\tau \approx 1.3$
\cite{richard2002surface,schutzius2015}.
The longer contact time for bouncing Leidenfrost droplets
may allow the thermocapillary-induced
flow to energize the droplet,
resulting in higher
stored surface energy
and subsequently
an unusually high
recovery of gravitational potential
of the bouncing motion.
This is best illustrated
by examining the
relation between the amplitude
$A_{k+1}$ of bouncing cycle $k+1$
and that of the immediately preceding cycle, $A_{k}$.
As shown in Fig.~\ref{fig4}c,
the amplitudes $A_{k}$
and $A_{k+1}$
of any two consecutive cycles
are always almost identical,
indicating that the
restitution coefficient
$\epsilon_ k =({A_{k+1}}/{A_{k}})^{1/2}$
of any cycle $k$ is approximately unity.
The average restitution coefficient $\epsilon$
for the entire bouncing
stage, shown in Fig.~\ref{fig4}c,
ranges from 0.97 to 0.99
with the surface temperature varying from 300$^\circ $C to 440$^\circ $C.
This is remarkably higher
than
the restitution coefficient
of droplets
impinging on suprehydrophobic surfaces
($0.5\le\epsilon\le$0.9) \cite{richard2000bouncing,schutzius2015},
or that on thin liquid films
($0.75\le\epsilon\le0.9$)
\cite{hao2015superhydrophobic}
(see Fig.~\ref{fig4}e).
We note that
$\epsilon \approx 1$
suggests that the energy lost to viscous dissipation
is almost completely compensated by
the kinetic energy gained
by the thermocapillary-induced flow.
The fact that $\epsilon$ is independent
of temperature
highlights a distinctive feature of
thermocapillary stress:
the liquid temperature
at the droplet's bottom surface
remains approximately fixed
at boiling temperature,
leading
to a stable vertical
temperature difference across the droplet and subsequently
a thermocapillary stress that depends weakly on
the temperature of the solid surface.

Leidenfrost droplets, therefore,
always
set off
to a series of trampolining motions with a variety of
rhythms.
A sufficiently large Leidenfrost droplet always starts
hovering on the heated surface
with fluctuating frequencies
until its size becomes comparable to the
liquid's capillary length, at which the droplet starts bobbing, i.e., oscillating
with more regular frequencies but without bouncing.
A transition from bobbing to bouncing occurs when the Froude number
becomes larger than unity, signifying that the
upward acceleration caused by internal flows overcomes the gravitational acceleration.
In the bouncing stage, the droplet's dynamics are driven by parametric
oscillation with occasional resonances whereby the oscillation frequency drops
and the amplitude trebles. The bouncing motions of the Leidenfrost
droplet is also characterized by an unusually high restitution coefficient resulted from
the thermocapillary-induced flow compensating the lost energy due to viscosity.
Our findings of the spontaneous trampolining motions of Leidenfrost droplets
complete their portraiture as an active system capable of creating its own motions
and energetic states. The underlying active mechanism may
provide a promising avenue for frictionless liquid manipulation and transport,
as well as a potential strategy for energy harvesting.

This research was conducted in collaboration with
HP Inc. and supported by Nanyang Technological University
and the Singapore Government through the
Industry Alignment Fund--Industry Collaboration Projects Grant.


%

\end{document}